# Slow Relaxation, Spatial Mobility Gradients and Vitrification in Confined Films

Stephen Mirigian[1,4] and Kenneth S. Schweizer[1-4,*]

Departments of Materials Science[1], Chemistry[2], Chemical and Biomolecular Engineering[3] and Frederick Seitz Materials Research Laboratory[4], University of Illinois, Urbana, IL 61801

*kschweiz@illinois.edu

**Abstract**

Two decades of experimental research indicates that spatial confinement of glass-forming molecular and polymeric liquids results in major changes of their slow dynamics beginning at large confinement distances. A fundamental understanding remains elusive given the generic complexity of activated relaxation in supercooled liquids and the major complications of geometric confinement, interfacial effects and spatial inhomogeneity. We construct a predictive, quantitative, force-level theory of relaxation in free-standing films for the central question of the nature of the spatial mobility gradient. The key new idea is that vapor interfaces speed up barrier hopping in two distinct, but coupled, ways by reducing near surface local caging constraints and spatially long range collective elastic distortion. Effective vitrification temperatures, dynamic length scales, and mobile layer thicknesses naturally follow. Our results provide a unified basis for central observations of dynamic and pseudo-thermodynamic measurements.

Glass forming liquids undergo remarkable changes of dynamics at rather large confinement distances; in some cases shifts of the apparent vitrification temperature commence at 25-50 nm or beyond for polymer films[1-5]. It has long been hoped this phenomenon holds critical clues about cooperative relaxation in bulk supercooled liquids[6], but this remains unrealized due to the strong effect of interfaces, e.g., vapor vs. solid, surface chemistry[2]. This problem is also crucial for diverse materials applications[7] and the formation of "ultra-stable" glasses[8]. A major mystery is the breakdown of accepted inter-relationships between different experimental probes (e.g., thermodynamic vs. dynamic) that hold in the bulk.[1,2,9,10]. However, it is widely acknowledged that the central question is the nature of a spatial gradient of mobility[2,11-13].

Time-dependent measurements are the most fundamental measure of glassy dynamics.

Recent experiments[14] on *free-standing* polymer thin films find a 2-step decay of a probe molecule reorientational correlation function, C(t), at ~10-30 K below the bulk glass transition temperature, $T_g$. This suggests a fast relaxing (by ~3-4 orders of magnitude) population of segments within several nm of the vapor interface corresponding to a temperature-dependent "mobile layer"[14,15], and a slow bulk-like population in the film interior. Not far above $T_g$ the 2-step decay seems to disappear with the slow and fast processes "merging". The generality of such phenomena is suggested by their near independence of polymer chemistry and chain length[14], a ~7 decade speed up of diffusion[16] at $T_g$ and viscous flow[17] of molecular glass-formers near the film surface, reduced interfacial viscosity[18], and other measurements[19]. Especially notable is the nanoparticle embedding measurements which directly detect a mobile surface layer.[15]

Theoretical progress has been modest[2,13,20,21] due to the inherent complexity of bulk activated relaxation, confinement and interfacial interactions. Simulations[2,13,22,23] provide valuable insights but cannot access the deeply supercooled regime since they probe only down to of order the dynamic crossover temperature[6,13], $T_c$. Here, the bulk relaxation time is ~8-10 orders of magnitude faster than at $T_g$, the mobility enhancement at a free surface is only ~3-4 orders of magnitude, and a 2-step form of C(t) is not observed[2,13,23]. In this Communication we construct a no adjustable parameter, force-level theory of the mobility gradient, and determine its multi-variant consequences in free-standing[2,3,24-26] films. The key idea is that vapor interfaces accelerate hopping by *both* reducing near surface local caging constraints and long range collective elastic distortion. Effective vitrification temperatures and length scales naturally follow. The focus is on dynamics, but contact is made with pseudo-thermodynamic measurements[1-3].

The enabling foundation for our work is the bulk "elastically collective nonlinear Langevin equation" (ECNLE) theory[27,28] based on the concept of a particle displacement, r(t), dependent microscopic *dynamic* free energy, $F_{dyn}(r)$, that quantifies the effective force on a moving particle due to its surroundings (see Figure 1). For a fluid of spheres (diameter, d), $F_{dyn}(r) = F_0(r) + F_{cage}(r)$, where $F_0(r) = -3k_B T \ln(r)$ quantifies the driving force for unbounded diffusion, and $F_{cage}(r)$ quantifies intermolecular constraints which favor spatial localization and solid-like behavior and can be *a priori* calculated[28] from knowledge of fluid density ρ and the radial distribution function, g(r), or structure factor. Key features of $F_{dyn}(r)$ include the barrier for *local* cage re-arrangement, $\Delta F_B$, transient localization length, $r_{loc}$, barrier location, $r_B$, and jump distance, $\Delta r \equiv r_B - r_{loc} \approx 0.2 - 0.4d$. For deeply supercooled liquids, as originally proposed phenomenologically by Dyre[29], activated hopping requires a small expansion of the nearest neighbor shell and harmonic elastic distortion of the surrounding medium, resulting in an additional, spatially *non-local*, collective barrier[27,28]:

$$\Delta F_{elastic}^{bulk} = \rho(K_0/2) \int_{r_{cage}}^{\infty} dr\, 4\pi r^2 u^2(r) \qquad (1)$$

The displacement field[29] outside the cage radius ($r_{cage} \approx 1.5d$) is $u(r) = \Delta r_{eff}(r_{cage}/r)^2$, the cage dilation scale is $\Delta r_{eff} \propto \Delta r^2 / r_{cage} \leq r_{loc}$, and the spring constant describing localization (and dynamic shear modulus) is $K_0 = 3k_B T / r_{loc}^2$. The elastic barrier, $\Delta F_{elastic}^{bulk} \approx 12\phi \Delta r_{eff}^2 r_{cage}^3 d^{-3} K_0$, plays the central role in the deeply supercooled regime, $\tau_\alpha > 10^{-7\pm1}s$, and involves long range (scale invariant) motion with 90% of its total value requiring cooperative displacements out to a distance ~13d from the cage center[28].

Bulk ECNLE theory is rendered quantitatively predictive for thermal liquids by mapping[28] molecules to an effective hard sphere fluid that exactly reproduces the equilibrium dimensionless density fluctuation amplitude or compressibility of the real system, $S_0(T) = \rho k_B T \kappa_T$. This yields a temperature and material-specific effective packing fraction, $\phi(T)$, which determines structure and dynamical constraints. The resultant theory accurately captures relaxation in van der Waals liquids (e.g., orthoterphenyl (OTP), trisnapthylbenezene (TNB)) over 14 decades in time[27,28].

An interface can locally modify density, compressibility and molecular orientation. Our hypothesis for free-standing films (consistent with lack of measurable density changes) is these are second order effects and are ignored here. Rather, we emphasize three *generic* physical mechanisms of how a free surface modifies the spatially nonlocal activated relaxation event (Fig.1): (i) a "surface" effect close to the interface associated with reduced caging constraints, (ii) a long range "confinement" effect mainly due to collective elastic physics, and (iii) strong coupling between (i) and (ii) via a spatial gradient of elastic stiffness and cage expansion amplitude.

Cage rearrangement occurs via relatively large amplitude hopping (plot "a" in Fig.1) with a barrier due to forces exerted by nearest neighbors[27,28]. Within a distance $r_{cage}$ from a free surface, caging forces are reduced due to missing neighbors and hence $F_{cage}(r) \to \gamma(z) F_{cage}^{(bulk)}(r)$, where an elementary geometric calculation yields the ratio of nearest neighbors a distance z from the surface relative to its bulk analog as: $\gamma(z) = 0.5 - 0.25(z/r_{cage})^3 \left[1 - 3(r_{cage}/z)^2\right]$ for $z \le r_{cage}$. For $z > r_{cage}$, $\gamma \to 1$ and the bulk $F_{dyn}(r)$ is recovered; when z→0, $\gamma \to 0.5$, corresponding to missing half of the nearest neighbors at the surface[21].

The elastic penalty associated with long range displacements is also weakened near the interface due to missing neighbors (plot "b" in Fig.1). This softening decreases continuously with distance from the surface, becoming bulk-like (plot "c") deep in the film, as reflected in the color gradient darkening in Fig.1. Since there are no particles outside the film, the free surface effectively "cuts off" part of the collective barrier. We treat this effect as a simple cut off of the elastic deformation field thereby yielding :

$$\Delta F_{elastic}(z) = (\rho/2) \int_V d\vec{r}\, u(\vec{r};z)^2 K_0(\vec{r}) \quad (2)$$

where $V$ is the film volume. The strain field now depends on *both* the distance from the cage center and the location of the relaxation event in the film. The mean time associated with activated barrier crossing follows as[28]

$$\frac{\tau_\alpha}{\tau_s} = 1 + \frac{2\pi\left(k_B T/d^2\right)}{\sqrt{K_0 K_B}} e^{(\Delta F_B + \Delta F_{elastic})/k_B T} \quad (3)$$

where $K_0$, the barrier curvature $K_B$, both barriers, and the alpha time *all* depend on location in the film and its thickness, z and h, respectively; $\tau_s$ is the known, non-activated, bulk, short time relaxation process time scale[28]. Calculation of the relaxation time function $\tau_\alpha(T,h,z)$ then allows the prediction of characteristic length scales and apparent vitrification temperatures relevant to diverse experiments.

We illustrate our central predictions using equilibrium $S_0(T)$ input[30] for polystyrene (PS) melts and the Kuhn length[31] ($l_K$, twice the persistence length) as the dynamically relevant coarse graining variable. This corresponds to a liquid of disconnected Kuhn spheres, in the spirit of the molecular liquid mapping[28]; for PS, $d=l_K\sim1.2$nm. All calculations are nearly identical for molecules such as OTP and TNB[28]. Spatial gradients of dynamical properties are a continuous function of location in the film, but one should keep in mind that experiments have a finite resolution, e.g., $\sim(0.5\text{-}1)d$.

The main frame of Fig.2 shows the local cage, long range elastic, and total barriers, as a function of nondimensionalized film location $\zeta = 1 - 2z/h$, each normalized by its bulk value. Results are for h=36 nm~30d at the bulk $T_g$, defined as when $\tau_\alpha = 100s$; the bulk barriers are $\Delta F_B \approx 14 k_B T$ and $\Delta F_{elastic} \approx 18 k_B T$. The local barrier is strongly reduced close to the surface and saturates to its bulk value a distance $r_{cage}$ into the film. In *qualitative* contrast, the elastic barrier, while strongly reduced near the surface, is suppressed far into the film as a unique consequence of the nonlocal[28,29] nature of the alpha process *and* its coupling to near surface cage weakening.

The inset of Fig.2 shows the corresponding relaxation time gradient at five temperatures that straddle the bulk $T_g$. The alpha time is massively faster near the surface, and varies weakly with temperature. On the scale of d, relaxation near the surface at the bulk $T_g$ speeds up by ~6-8 orders of magnitude, consistent with (near) surface diffusion in molecular systems[16]. The calculation at 426 K mimics the dynamic crossover temperature, $T_c\sim1.2 T_g$, where the collective barrier is of minor importance. This is the regime probed in simulations, and the long tail into the film is largely absent since elastic distortion is very weak, and the gradient covers only ~3 orders of magnitude.

The calculations in Fig.2 allow a film-averaged relaxation function, $C(t) = \langle e^{-t/\tau_\alpha(z)} \rangle_z$, to be computed, as shown in Figure 3. In a temperature window modestly below $T_g$, and a time window germane to experiment[14], C(t) decays in two steps both of which are nonexponential solely due to the spatial mobility gradient. The fast, surface-related process has a rather low amplitude of ~15%, while the slower bulk-like process has an amplitude ~80%. The fast process is of a highly stretched (KWW) form, $\sim e^{-(t/\tau_{KWW})^{\beta_K}}$, where $\beta_K \approx 0.5 \to 0.27$ for T=376→346 K. In contrast, the slow process exhibits a much larger and far less temperature-dependent $\beta_K \approx 0.9 \to 0.8$. The inset of Fig.3 shows the fast process time grows more slowly with cooling, becoming ~2 orders of magnitude shorter than the slow process. Far enough above the bulk $T_g$, a two-KWW fit is not sensible since the mobility gradient is much weaker (per in simulations[13,22,23]) and the fast process falls outside the experimental time window[14]. We estimate via extrapolation a "merging" temperature at ~$T_g$+25K; experiments find ~$T_g$+15K.

The inset of Fig. 3 also shows a mobile liquid-like layer thickness, $z^*$, defined as the part of the film that relaxes faster than 100s. We find, e.g., $z^* \sim$ 2-3 nm at 5K below the bulk $T_g$, and results for different film thicknesses essentially overlap except at $T_g$ where (by definition) $z^* \to$ h/2. A measureable mobile layer is predicted to be undetectable 30-40 K below $T_g$ *if* the length scale resolution is $\leq d$. Other experiments with different resolutions[15,19] report mobile layers 50-80K below $T_g$, which is not inconsistent with our results in Fig.3.

All calculations in Figs. 2 and 3 are in qualitative accord with recent probe rotation measurements[14]. But there are quantitative deviations, e.g., the degree of stretching, and difference between the fast and slow relaxation times (2 vs. ~3-4 orders of magnitude), are smaller than observed. This is perhaps unsurprising given the theory and molecular model are approximate, experiments measure probe (not matrix) dynamics, the space-time resolution issue, and the existence of a relaxation time distribution in the bulk ignored here for simplicity.

We now consider the subtle question of an effective glass transition temperature[1,2,32]. The main frame of Figure 4 shows a purely dynamical $T_g$ defined as when the film averaged relaxation time reaches 100s, $\langle \tau_\alpha(z) \rangle_h |_{T_g} = 100\,\text{s}$ (corresponding, e.g., to a dielectric loss inverse peak frequency), and a "thermodynamic-like" alternative, $\langle T_g(z) \rangle_h$, where $\tau_\alpha(T_g(z;h)) \equiv 100s$; representative $T_g$-gradients are shown in the inset. The $\langle T_g(z) \rangle_h$ results show a significantly larger $T_g$ drop than the purely dynamic analog as a consequence of how the mobility gradient is averaged. They are reminiscent of pseudo-thermodynamic measurements; e.g., at h=10 nm, $\langle T_g(z) \rangle_h$ decreases by ~20K, consistent with ellipsometry experiments for ~10 nm low and moderate molecular weight free-standing PS films[3,24] that find a ~25K $T_g$ reduction. The thickness dependences are well fit (solid curves) by the empirical form that describes various experiments and simulations[13], $T_g(h) = T_{g,bulk}(1+\xi/h)^{-1}$, with $\xi < d$.

To make direct contact with ellipsometry data we construct a thermodynamic *effective* 2-layer model for the thermal expansivity per ref.[32]. Using the computed mobile layer thickness, $z^*$, one has $\alpha_{eff}(T) = \alpha_g(1-2z^*(T)/h) + \alpha_l 2z^*(T)/h$, where the liquid (glass) $\alpha_l = .0004\text{K}^{-1}$ ($\alpha_g = .0001\text{K}^{-1}$). Taking the film thickness at a low $T_0$ to be $h_0$, one has $h(T)/h_0 = 1 + \int_{T_0}^{T} dT' \alpha_{eff}(T')$. Representative calculations (inset of Fig.4) show the key experimental features[1-3] are captured, including decreasing (increasing) contrast (breadth) of the *apparent* liquid→glass transition as the film thins. The $T_g$ values determined from the intersection of linear fits to the high and low temperature regimes are shown as open circles in the main frame of Fig. 4. Very interestingly, they agree essentially exactly with our $\langle T_g(z) \rangle_h$ calculations. These results provide new insights concerning the connection between an apparent $T_g$ determined by falling out of thermodynamic equilibrium and one deduced based on equilibrated dynamics. The ideas[32] that the ellipsometric $T_g$ is "some kind of average of the gradient of $\tau_\alpha$", the kink in h(T) does not indicate a real thermodynamic glass transition but rather reflects a mobile layer, and the dilatometric $T_g$ is a convolution of enhanced surface mobility and a dynamical penetration length, all find theoretical support in our work.

In conclusion, we have constructed a quantitative, force level theory for how confinement in free-standing thin films introduces a mobility gradient as encoded in $\tau_\alpha(T,z,h)$. Diverse consequences appear consistent with experiment, and the theory has demonstrated[27,28] material-specific predictive power. Of course, much remains to be done, including incorporating anisotropic corrections to our simple cut off model of a radially symmetric elastic deformation field. Nonetheless, the present approach provides a foundation to treat diverse phenomena such as puzzling influences of chemistry on $T_g$ shifts[33], explicit effects of polymer connectivity, the consequences of solid surfaces, mechanical properties[34], and non-planar geometries such as spherical droplets[35]. Work in all these directions is in progress.

**Acknowledgement.** This work was supported by the U.S. Department of Energy, Basic Energy Sciences, Materials Science Division via Oak Ridge National Laboratory. Informative discussions with Mark Ediger, Zahra Fakhraai and Greg McKenna are gratefully acknowledged.

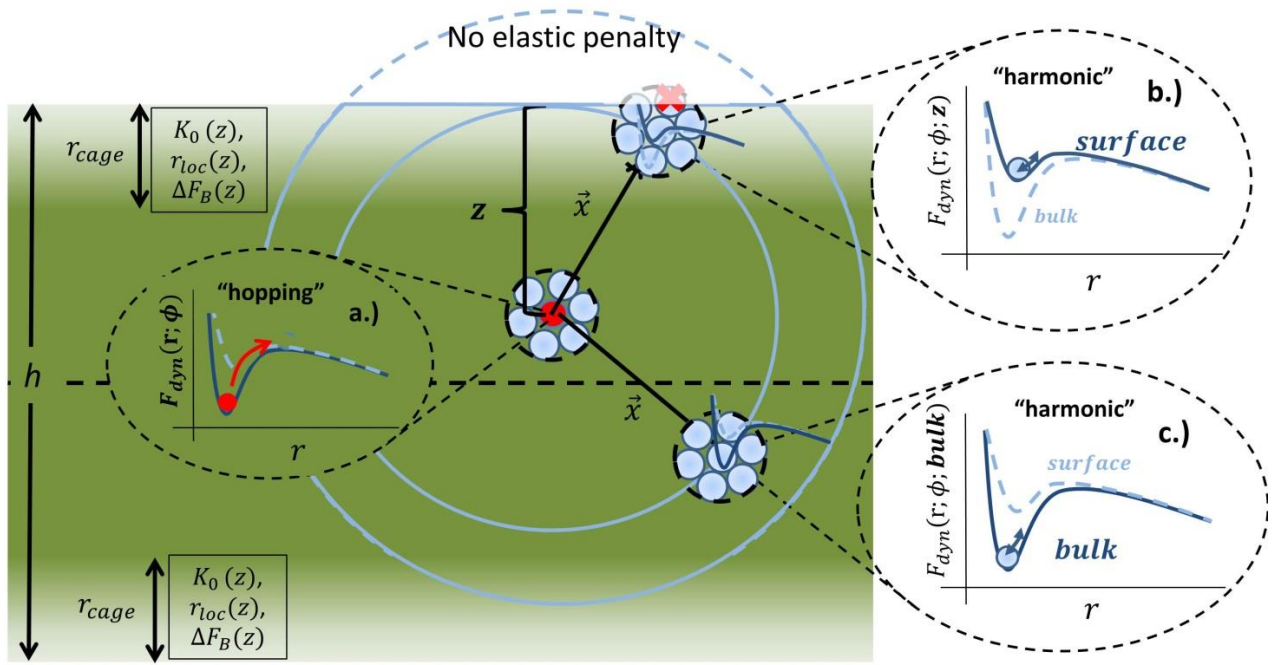

Fig. 1. Conceptual schematic of dynamical processes and key dynamic free energy features. The film thickness is h and the distance of a local rearrangement event from the film surface is z.  a.) Hopping requires surmounting a local barrier.  b.) Particles near the surface experience a reduced caging force due to missing neighbors, resulting in a film-location-dependent softer confinement potential. c.) Particles far from a free surface experience the bulk dynamic free energy.  The long range elastic barrier is a sum of the elastic energy penalty for harmonic motion throughout the spatially heterogeneous film.

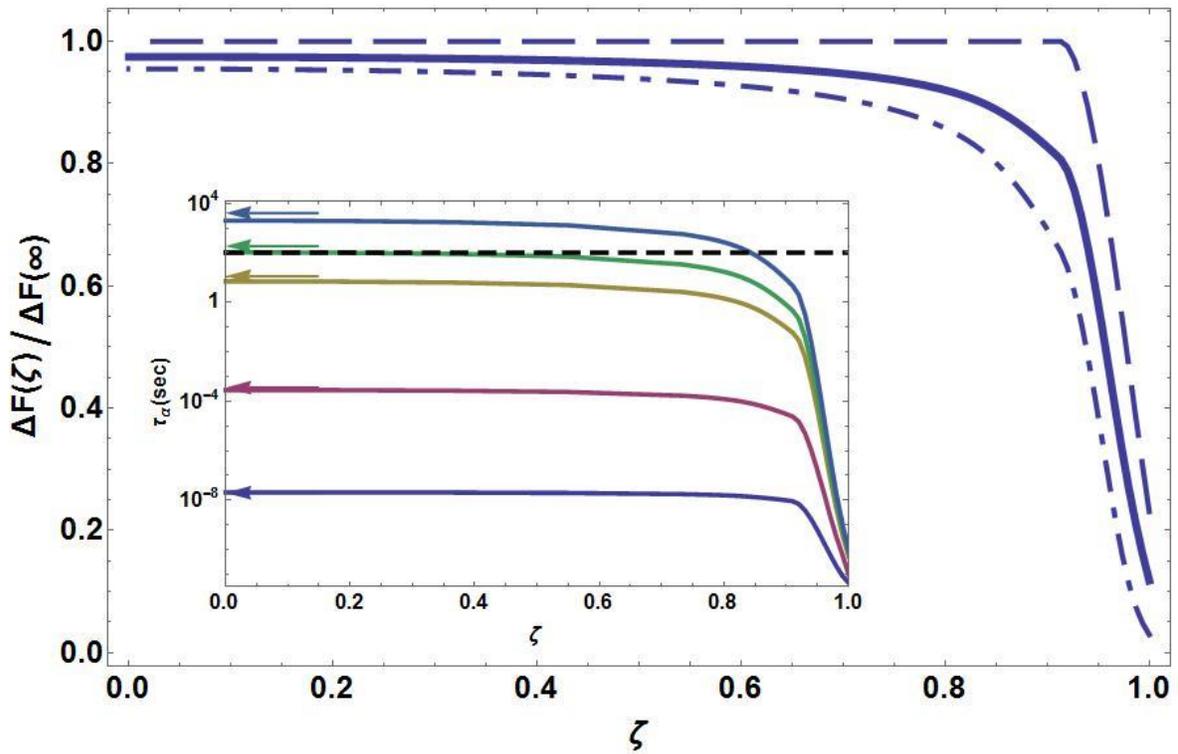

Fig 2. Local cage (dashed), elastic (dash-dot), and total (solid) barriers for a film thickness h=36 nm at the bulk $T_g$ as a function of reduced location, $\zeta = 1 - 2z/h$; each barrier is normalized by its bulk value. Inset: Corresponding relaxation time profiles at 426K (blue, near the bulk $T_c$), 386K (red), 361K (yellow), 356K (green, predicted bulk $T_g$), and 351K (gray). The horizontal black dashed line indicates kinetic vitrification, and arrows along the vertical axis indicate the bulk alpha time.

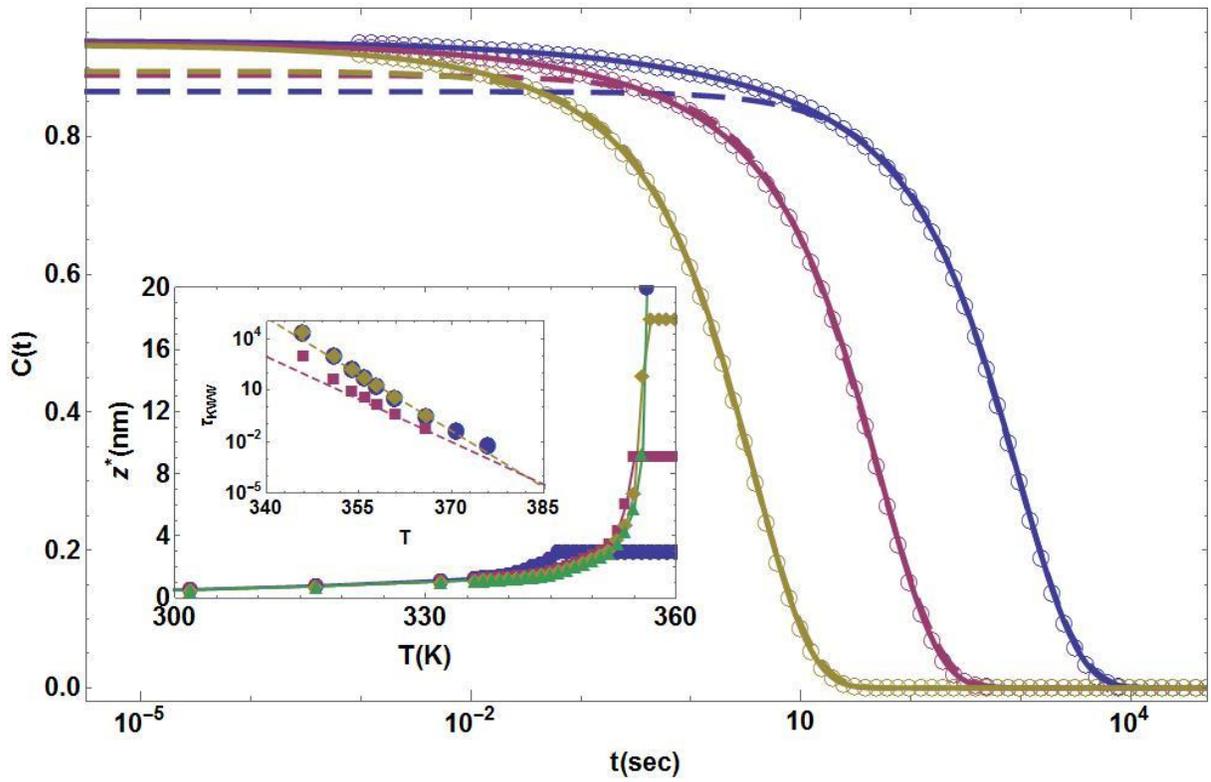

Fig. 3. Relaxation function for h=36 nm at T = 351K (blue circles), 356K (red circles, bulk $T_g$) and 361K (yellow circles). The solid (dashed) curves are double (single) KWW fits. Insets: Corresponding extracted relaxation times with an apparent merging point determined via extrapolation. Mobile layer thickness ($z^*$ in nm) as a function of temperature for 6nm (blue), 18nm (red), 36nm (yellow) and 180nm (green) films.

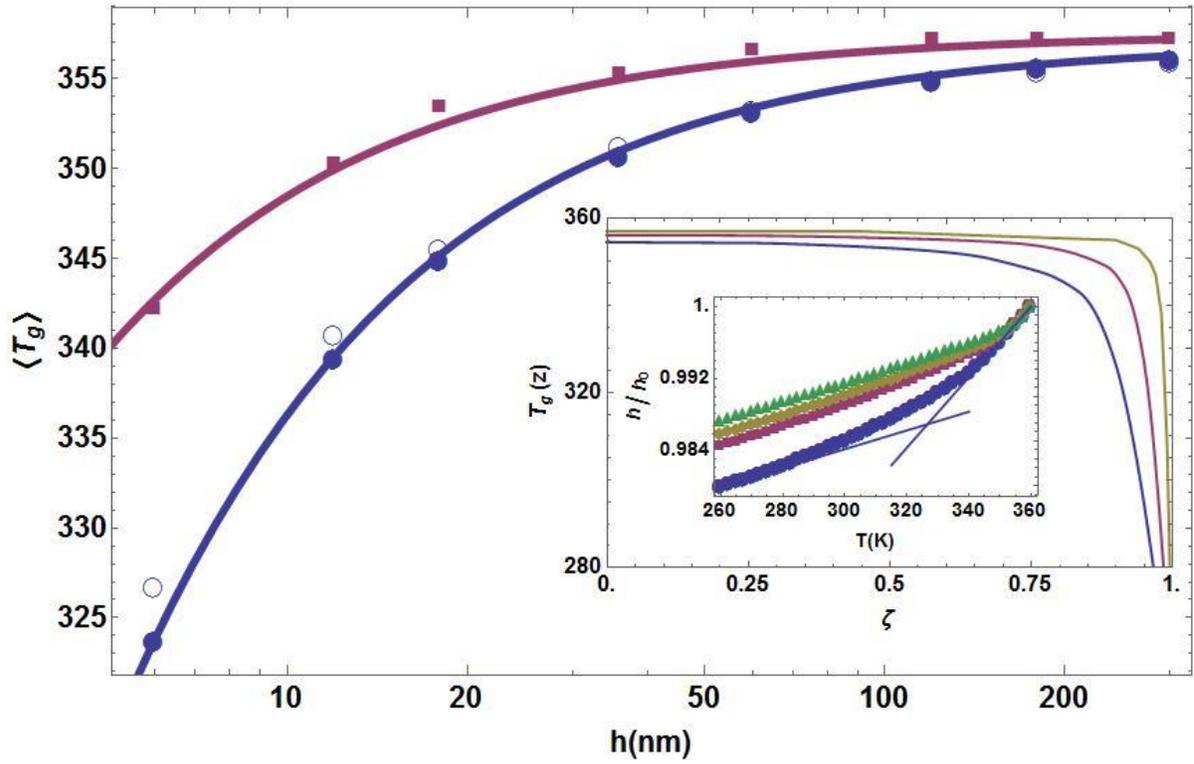

Fig 4. Film-averaged glass transition temperatures as a function of thickness: $\langle T_g(z)\rangle_h$ using the vitrification profile of the inset (closed circles), $T_g$ based on the dynamic criterion $\langle \tau_\alpha(z)\rangle_h|_{T_g} = 100\,\text{s}$ (red squares), and thermodynamic ellipsometric result (open circles) based on the h(T) calculations of the inset. Insets: Kinetic $T_g$ profile for film thicknesses of 18nm (blue), 36nm (red) and 120nm (yellow). Temperature variation of the film thickness for the same systems in the inset of Fig.3: 6nm (blue), 18nm (red), 36nm (yellow) and 180nm (green) films; extraction of a $T_g$ as measured using ellipsometry is indicated.